\documentclass[prl,twocolumn,showpacs,amsmath,amssymb,superscriptaddress]{revtex4}

\usepackage{graphicx}  % Include figure files
\usepackage{dcolumn}   % Align table columns on decimal point
\usepackage{bm}        % bold math
\usepackage[latin1]{inputenc}

\begin{document}
\preprint{v 1.03}

\title{Observation of sub-Fourier resonances in a quantum-chaotic system}

\author{Pascal Szriftgiser}
\author{Jean Ringot}
\affiliation{Laboratoire de Physique des Lasers, Atomes et
  Mol{\'e}cules, UMR CNRS 8523, Centre d'Etudes et de Recherches Laser et Applications,
Universit\'{e} des Sciences et Technologies de Lille, F-59655
Villeneuve d'Ascq Cedex, France}
\homepage{http://www.phlam.univ-lille1.fr}
\author{Dominique Delande}
\affiliation{Laboratoire Kastler-Brossel, Tour 12, Etage 1, 
Université Pierre et Marie Curie, 4 Place Jussieu, 
F-75252 Paris Cedex 05, France }
\author{Jean Claude Garreau}
\affiliation{Laboratoire de Physique des Lasers, Atomes et
  Mol{\'e}cules, UMR CNRS 8523, Centre d'Etudes et de Recherches Laser
  et Applications,
Universit\'{e} des Sciences et Technologies de Lille, F-59655
Villeneuve d'Ascq Cedex, France}
\homepage{http://www.phlam.univ-lille1.fr}

\date{\today}

\begin{abstract}
  We experimentally show  that the response
  of a quantum-chaotic  system can display resonance lines sharper
  than the inverse of the excitation duration.
  This allows us to discriminate two
  neighboring frequencies with a resolution nearly 40 times better
  than the limit set by the Fourier inequality. Furthermore, numerical
  studies indicate that there is no limit, but the loss of signal, to
  this resolution, opening ways for the development of sub-Fourier
  quantum-chaotic signal processing.
\end{abstract}

%\keywords{}
\maketitle

Any time-dependent signal can be characterized by its frequency spectrum, 
obtained by a Fourier transform. The Fourier inequality implies that
getting a narrow frequency spectrum requires that the corresponding
temporal signal lasts a long time. This inequality links the width
$\Delta f$ of the frequency spectrum and the temporal width 
$\Delta t$ of the signal \cite{note:Fourier} : $\Delta f \Delta t \ge
1/2$. A rule that is usually deduced from the above one states that 
two frequencies cannot be distinguished before a time proportional to the
inverse of their difference. Of course, if the signal/noise ratio 
is infinite, two
unresolved frequencies can still be distinguished using a good fit. This
is not the problem we are here interested in. We rather concentrate 
on the raw width of a resonance signal, before any fit is performed.
Physically, the Fourier
inequality sets a limit to the minimum width of a 
resonance line to the inverse of the time duration of the experiment,
a rule that applies also to quantum systems. For
instance, the width of Ramsey fringes \cite{ref:Ramsey} is limited by
the  time interval between the crossing times corresponding to the two Ramsey
zones, and the frequency width of an atomic or molecular resonance
absorption line is generally given by the inverse of the lifetime of
the excited state. However, these conclusions rely on the linear
relation between the system's response to the excitation, and
this limit can in principle be overcome provided
a suitable non-linearity is introduced. The simplest example is to use
a multiphoton resonance. Indeed, the frequency width of a $n$-photon 
excitation line is divided by a factor up to $n$ compared to a
single photon excitation, as recently observed with multiphoton Raman
transitions on atomic rubidium \cite{ref:Multiphoton}. In this case,
the sub-Fourier character originates from the fact that it is
$n^{th}$ harmonic of the external driving frequency that has to
be compared to the atomic frequency, and not the driving frequency
itself. As a consequence, such a device
is not very flexible, as it allows sub-Fourier lines only
at subharmonics of the intrinsic atomic frequency.
In the present 
paper, we show that, for a quantum system displaying chaos (which
implies the presence of an intrinsic nonlinearity), sub-Fourier
resonances can be widely observed, without the need of a resonance
with some internal frequency of the system.
They rely on a different process: the high sensitivity of a quantum chaotic
device to periodicity breaking.

The system we studied is an atomic version of the quasiperiodically
driven quantum kicked rotor described in \cite{ref:Bicolor}.
The periodic quantum kicked rotor \cite{ref:aQKR}
is a paradigmatic model for studies of time-dependent quantum-chaotic
systems. Its atomic version has been implemented in recent experiments
by many groups \cite{ref:Bicolor,ref:eQKR,ref:Raizen98},
and consists in placing
a laser-cooled atomic cloud in a periodically pulsed, far-detuned
($\Delta_L \approx$ 9.2 GHz $\approx 1.7 \times 10^3 \Gamma$ in the present case),
standing laser wave (SW). The atoms and the
electromagnetic field then interact via the so-called ``dipole''
force \cite{ref:Dipole}, which is conservative; the high laser-atom
detuning insuring that dissipative effects due to spontaneous emission
are negligible. The corresponding hamiltonian for a single atom is
then, in convenient dimensionless units \cite{ref:Raizen98}:
\begin{equation}
  H_1 = \frac{P^2}{2} + K \cos \theta \sum_{n=1}^{N_1} \delta_{\tau}(t-n)
  \label{eq:H1}
\end{equation}
where time is measured in units of the kick period $T_1 = 1/f_1$, $P$ is
the reduced momentum in units of $M/(2k_LT_1)$ ($k_L$ is the laser
wavenumber and $M$ the mass of the atom), $\theta=2k_Lz$ the reduced
position of the atom along the SW axis,
$K=\Omega^2T_1 \tau \hbar k_L^2/(2 M \Delta_L)$
the normalized kick intensity ($\Omega$ is the resonant Rabi
frequency of the SW beams), $N_1$ the number of kicks,
and $\delta_\tau (t)$ a Dirac-like
square function of width $\tau$ with $\tau \ll T_1$ [$\delta_\tau(t) =
1/\tau$ for $|t| \le \tau/2$ and $\delta_\tau(t)=0$ elsewhere]. In the
limit $\tau \rightarrow 0$, the dynamics of the system is
entirely determined by two parameters: the kick intensity (or {\it
stochasticity parameter}) $K$ and the effective Planck constant
$\hbar_{\mathrm{eff}} = 4 k_L^2T_1\hbar/M$. For $K \gg 1$,
the classical dynamics associated 
with the hamiltonian Eq.~(\ref{eq:H1}) is a chaotic, although perfectly
deterministic, ergodic diffusion in phase space, and the mean kinetic
energy $\langle P^2\rangle/2$ roughly grows linearly with time. Its
quantum dynamics is however completely different, and displays a
phenomenon known as ``dynamical localization" (DL), which is a
signature of quantum chaos \cite{ref:DL}. DL consists in the
suppression of the classical momentum diffusion (or a freezing of
the wave packet evolution) after some localization time $t_L$, due to
quantum destructive interferences among the various chaotic
classical trajectories. For $t > t_L$, the average momentum
distribution of the atom presents a characteristic time-independent
exponential shape $p(P,t > t_L) \sim \exp(-|P|/L)$, with a localization
length in the momentum space given by $L$. DL is intimately related to the 
periodic character of the kick sequence, and bears a close analogy
with Anderson's localization in disordered solids
\cite{ref:AndersonLoc}.

The sensitivity of DL to deviations from exact periodicity is at the
heart of the present experiment. Consider the following hamiltonian,
consisting in two series of pulses of frequencies $f_1$ and $f_2$,
with a ratio $r = f_2/f_1 \approx 1$, and relative phase $\varphi$:
\begin{equation}
  H_2 = H_1 + K \cos \theta \sum_{n=1}^{N_2<rN_1}
  \delta_\tau \big(t-\frac{n+\varphi/2\pi}{r}\big)
  \label{eq:H2}
\end{equation}
with the same normalizations as in Eq.~(\ref{eq:H1}).
If $r=1$ (or, more generally, equal to any rational number), the
perturbation is periodic and $H_2$ leads to dynamical localization. If
the value of $r$ is slightly shifted from 1 and turns irrational,
the kick series loses its temporal periodicity, DL disappears and the 
quantum evolution is on the average a continuous chaotic diffusion
leading to arbitrarily large momenta. The dependence of the DL
on the rational or irrational character of $r$ was recently observed
experimentally \cite{ref:Bicolor}. A fundamental question is: how
fast can the quantum system distinguish between a true simple rational
value of $r$ and a neighboring irrational value? The present
experiment gives an unexpected response to this question.

%FIGURE 1
\begin{figure}
\includegraphics[width=8cm,clip=]{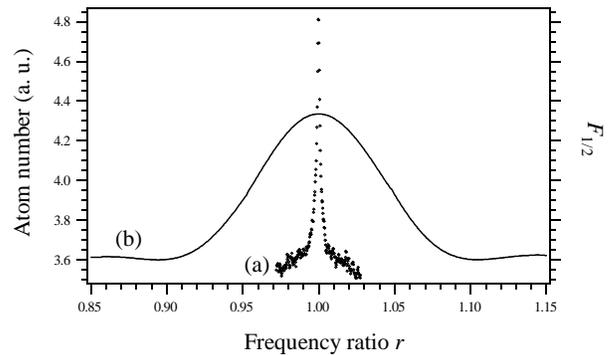} 
  \caption{\label{fig:SubF}
Below the Fourier limit. (a) Experimental measurement of the zero 
velocity atom number, $p(0)$, as a function of the ratio $r=f_2/f_1$ of the
two excitation frequencies, for the following
values of the parameters: $f_1 = 18$ kHz, $K = 42$, $N_1=10$. The pulse 
duration is $\tau =3$ $\mu$s and in order to avoid pulses overlap we
set $\varphi= \pi$. Averaging: 100 times. (b) $F_{1/2}(r)$, allowing
comparison with the Fourier transform of the kick sequence, this curve
is drawn with an arbitrary amplitude and offset.}
\end{figure}

In the experiment, a sample of cold cesium atoms issued from a
standard magneto-optical trap is kicked by a far-detuned stationary
wave (SW) with two series of kicks of frequencies $f_1$ and $f_2$ for
a duration $T = N_1T_1$, according to the hamiltonian
Eq.~(\ref{eq:H2}). The SW pulses are assumed to be short enough that 
the atomic motion during a pulse can be neglected, and the SW acts,
with respect to atomic center of mass dynamics, as an instantaneous 
kick. Just after the end of the kick sequence, the number of zero
momentum atoms $p(P=0)$ is probed with a Raman velocity selective
excitation \cite{ref:Raman}. Since the atom-number is conserved during
the kick sequence, $p(0)$ is proportional to the inverse of the typical 
momentum at the end of the pulse sequence, i.e. $1/\sqrt{\langle P^2 \rangle}$,
and thus measures the degree of localization: the wider the momentum
distribution, the smaller the value of $p(0)$. Whereas $f_1$ is kept
constant, this procedure is repeated for various values of $f_2 =
rf_1$, leading to the resonance line shown in
Fig.~\ref{fig:SubF}. As expected, this line has a maximum at 
$r = 1$ where DL is present. A remarkable feature is that its 
width is extremely narrow and easily beats the standard Fourier
limitation, with a $\Delta r_{\mathrm{exp}} = 0.0026$, ($\Delta$ is the
full width at half maximum, FWHM) and thus $\Delta f_2 = f_1 \Delta
r_{\mathrm{exp}} \approx 47$ Hz, yielding:
\begin{equation}
  \Delta f_2 T \approx \frac{1}{38} \ll 1
  \label{eq:SubF}
\end{equation}

\begin{figure}
\includegraphics[width=8cm,clip=]{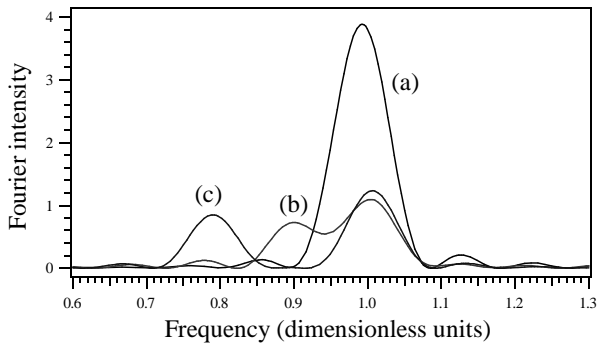} 
  \caption{\label{fig:F12}
First harmonic of the Fourier transform of the kick sequence. The 
parameters are: $T = 555$ $\mu$s, and $f_1 = 18$ kHz. (a) For $r = 0.98$
the two frequencies are not resolved; (b) $r = 0.93$ the two frequencies
are barely resolved; (c) $r = 0.80$, the two frequencies are clearly
resolved. Note that the experimental resolution is $\Delta r_{\mathrm{exp}} =
0.0026$, the frequency are thus well distinguished for  $r \not\in [0.9987,1.0013]$.}
\end{figure}

Indeed, in a first approach, one would consider the frequency spectrum
of the excitation formed by the two series of pulses. The Fourier
spectrum of an infinite sequence of pulses of frequency $f_1$ is a
comb of peaks separated by $f_1$. Since the sequence has a finite
duration $T$, each peak in the Fourier transform has a width $1/T$.
Furthermore, since the pulse duration $\tau$
is also finite, the whole spectrum is multiplied by the factor:
\begin{equation}
  \left( \frac{\sin(\pi f_1 \tau)}{\pi f_1 \tau} \right)^2 \text{ .}
  \label{eq:Sinctau}
\end{equation}
When two sequences with slightly different frequencies are combined, the 
resulting Fourier spectrum has two series of peaks that may overlap.
This is visible in Fig. \ref{fig:F12}, which shows, for
different values of $r$, the squared modulus of the Fourier transform
of the double kick sequence (with $\varphi = 0$), in the neighborhood
of the fundamental frequency $f_1$. Consider the function
$F_{1/2}(r)$, defined as the squared modulus of the Fourier transform
at frequency $f_{1/2} = (f_1+f_2)/2 = f_1 (1+r)/2$. This function is large 
when the two peaks overlap and goes to zero when the two frequencies
are completely resolved, giving a quantitative measure of the Fourier
resolution. In order to make a comparison with the experimental data,
$F_{1/2}(r)$ is also plotted in Fig.~\ref{fig:SubF}. Its width
$\Delta F_{1/2}(r)$ is 0.091, and $\Delta F_{1/2}(r)/\Delta r_{\mathrm{exp}}
\approx 35$, clearly showing that Fourier limit has been overcome. 

%FIGURE 2

\begin{figure}
\includegraphics[width=8cm,clip=]{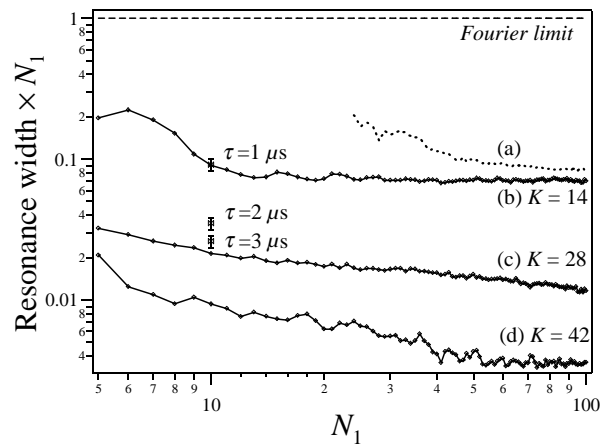} 
  \caption{\label{fig:Ndep}
Temporal evolution of the width of the resonance. The numerically
calculated width is multiplied by the number of kicks $N_1$ and
plotted as a function of $N_1$, the total excitation time in
normalized units. The Fourier limit thus corresponds to the horizontal
line of ordinate 1. The solid lines (b), (c) and (d)
correspond to three different
values of the parameter $K$. The displayed experimental points correspond to
three different values of the pulse duration $\tau =$ 1, 2 and 3
$\mu$s. The dotted line (a) corresponds to the numerical simulation using
the modulated series described by Eq.~(\protect \ref{eq:Mod}).}
\end{figure}

If the width of the resonance line is not Fourier-limited, what is the
origin of the process giving such a resolution? A first idea coming to
mind is that each atom is not excited just by the fundamental
frequencies $f_1$ and $f_2$, but also by their harmonics. The 
$j^{th}$ harmonic of each fundamental frequency has the same
width $1/ T$, but their frequency separation $j(f_1 - f_2)$ is
increased by a factor $j$. One would thus attribute the 
high resolution of the system to the presence of the high
harmonics in the excitation spectrum itself. In fact, this interpretation
is invalid for several reasons. Firstly, the 
experimental kicks are not real $\delta$-peaks, but have a finite
duration $\tau$ which implies that the weights of the high harmonics
are small. For instance, with $\tau=3$ $\mu$s, as for the curve shown in
Fig.~\ref{fig:SubF}, only harmonics up to $j=18$ lie in the
central lobe of function~(\ref{eq:Sinctau}). The harmonic
$j = 35$, whose frequency separation would corresponds to the experimental
resolution observed in Fig.~\ref{fig:SubF}, lies beyond the first lobe with a
very small intensity (less than 2\% of the intensity of the first
harmonic). Secondly, when $\tau$ is increased, with a constant pulse height 
(which implies that $K$ is
increased, as it is proportional to $\tau$), the relative weight of the high 
harmonics {\em decreases}, because of the factor Eq.~(\ref{eq:Sinctau}). But we
verified both experimentally and numerically that the system's
resolution {\em increases}: the resonance lines becomes narrower
(see Fig.~\ref{fig:Ndep}). Finally, we also performed numerical simulations using a
slightly different hamiltonian, in which the two-frequency excitation
is given by:
\begin{equation}
  \left[1+A \cos(2\pi rt)\right] \sum_{n=1}^N \delta_{\tau}(t-n) \text{ .}
  \label{eq:Mod}
\end{equation}
In this case, the second frequency is introduced as a
modulation of the intensity of the pulses. The spectrum of this
excitation can be calculated straightforwardly, and is composed of
series of peaks at frequencies $\omega_j/2\pi = j$ ($j$ is an
integer) with related sidebands at $\omega/2 \pi =j \pm (r-1)$. The frequency
separation between the two series of peaks in
the spectrum is $r-1$, independent of $j$: the high harmonics do not
provide higher resolution. Despite that, the numerical simulation
shown in Fig.~\ref{fig:Ndep} is clearly below the Fourier limit.
The above arguments thus neatly rule out the effect of high
harmonics as a dominant process at work in our experiment.
The observed ultra-narrow resonance line is in fact due to the highly
sensitive response of the quantum chaotic system itself to the
excitation, as we will no explain with semi-heuristic elements.

%FIGURE 3

In the experiment, the atomic wave packet undergoes a series
of kicks separated by a free evolution. In momentum representation,
a free evolution during a time $t$ amounts in 
multiplying the wave function by a phase factor $\exp(-i\Phi)$ with
$\Phi = P^2 t/(2\hbar_{\mathrm{eff}})$. For simplicity, consider the particular
case $\varphi = 0$, without loss of generality. In this case, the time
interval between the last kick ($N^{th}$ kick) of the first
series and the corresponding kick of the second the series is $N
(r-1)/r$. The system can resolve the two frequencies if 
the phase evolution $\Phi$ during this interval, is of the order of one: 
\begin{equation}
  \Phi(N) = \frac{\langle P^2 \rangle}{2\hbar_{\mathrm{eff}}} N
  \frac{r-1}{r} \approx 1 \text{ .}
  \label{eq:PhiThr}
\end{equation}
Before applying Eq.~(\ref{eq:PhiThr}), it is important to distinguish
two different regimes: (a) before the localization time, $N < N_L$
(where $N_L = t_L/T_1$), the dynamics is a classical diffusion, and
$\langle P^2 \rangle \propto N$; (b) for $N > N_L$ the momentum
distribution is localized and $\langle P^2 \rangle$ is
roughly constant. Thus, according to Eq.~(\ref{eq:PhiThr}), the resolution
$(r-1)$ has a fast, unusual, decrease as $N^{-2}$ as long as $N<N_L$,
and a slower decrease as $N^{-1}$ for $N>N_L$. Otherwise stated,
for times $t < t_L$ the system takes an advance over the Fourier
limit that it preserves when $t > t_L$. Effectively, the numerical
simulations in Fig.~\ref{fig:Ndep} indicate that the linewidth
decreases faster than the standard Fourier decrease, as $N^{-1}$. The higher
resolution results, on the one hand, from the sensitivity of the
quantum interference process to frequency differences, and, on the 
other hand, from the amplification of such sensitivity by a
chaotic diffusion. The underlying physical process relies
on long-range correlations in the momentum space. Breaking exact
time periodicity of the excitation destroys those correlations and,
as a consequence, the localization. This also suggests that the
frequency resolution improves as the dynamics becomes more chaotic,
as $K$ increases. Numerically, no lower limit has been found to the
resolution, which can reach several orders of magnitude below the
Fourier limit. There are nevertheless limits for real applications
since the signal $p(P=0)$ decreases with increasing $K$, roughly like
$1/K$. In our experiment, the signal/noise ratio is of the order of
100, which in principle allows widths below 1/100 of the Fourier
limit. There is however an experimental imperfection:  the SW waist is
only about 1.6 times the size of the atomic cloud: depending on their
position in the laser beam, the atoms experience different
values of $K$. This leads to an inhomogeneous broadening of the
resonance line, which explains the discrepancies between experimental
data and the numerical simulation in Fig.~\ref{fig:Ndep}. Improvements
towards narrower sub-Fourier lines are currently under study.

A few additional remarks might be done. Firstly, sub-Fourier behavior 
is allowed because the resonance lines are not the direct Fourier transform of a temporal 
signal. Secondly, the observation of sub-Fourier lines is not limited to a specific 
choice of parameters. The curve displayed in Fig.~\ref{fig:SubF} is the 
most sub-Fourier one (by a factor 35) that we have observed,
but sub-Fourier widths are a standard behavior. Thirdly,
since the process described here
involves two frequencies, it is quite analogous to a frequency
measurement by the heterodyning technique. This might be exploited
for realizing ultra-fast frequency locking to a standard frequency,
using $p(0)$ as an error signal, that would thus respond in a
sub-Fourier time. In a completely different field, an analogy can be
made between the present experiment and near field optical microscopy
where details of size far below the diffraction limit can be resolved.
In both cases, the resolution is obtained at the price of a signal loss.

In conclusion, we have experimentally demonstrated the possibility of
using of a quantum-chaotic device to discriminate two
frequencies, far below the Fourier limit. The principle relies on the
sensitivity of quantum interferences to periodicity breaking,
amplified by a chaotic diffusive process. No physical law has been
broken, but the very widely used ``rule of thumb" stating that the
minimum width of a quantum resonance should be limited by the excitation 
time is here not valid, opening the way to a new field of quantum-chaotic 
signal processing.

\begin{acknowledgments}
Laboratoire de Physique des
Lasers, Atomes et Molécules (PhLAM) is Unité Mixte de Recherche
UMR 8523 du CNRS et de l'Université des Sciences et Technologies de
Lille. Centre d'Etudes et de Recherches Laser et Applications (CERLA)
is supported by Ministère de la Recherche, Région Nord-Pas de Calais
and Fonds Européen de Développement Economique des Régions (FEDER).
Laboratoire Kastler-Brossel de l'Université Pierre et Marie Curie et
de l'Ecole Normale Supérieure is UMR 8552 du CNRS. CPU time on a NEC
SX5 computer has been provided by IDRIS.
\end{acknowledgments}


\begin{thebibliography}{99}
\bibitem{note:Fourier} Technically, $\Delta f$ refers to the square root of
  the variance of the frequency distribution, while $\Delta t$ is the square
  root of the variance of the temporal distribution of the signal. For
  most signal shapes, they can be associated with the width
  (at half maximum) of the frequency signal and its time duration,
  respectively.
\end{thebibliography}

\end{document}